\documentstyle[twocolumn,psfig]{article}

\def\ah{a_{\rm h}}
\def\alc{a_{\rm lc}}

\def\rc{r_{\rm c}}
\def\rcrit{r_{\rm crit}}
\def\rD{r_{\rm D}}
\def\rw{r_{\rm w}}
\def\thetalc{\theta_{\rm lc}}
\def\vej{v_{\rm ej}}
\def\vr{\vec r}
\def\vv{\vec v}
\def\vR{{\vec R}}
\def\vsoft{v_\epsilon}
\def\Fcm{F_{\rm cm}}
\def\M12{M_{12}}
\def\Mej{M_{\rm ej}}
\def\Vbin{V_{\rm bin}}

\def\BH{{\sc bh}}
\def\BHs{{\sc bh}s}
\def\mbh{m_{\rm bh}}
\def\soft{\epsilon}

\def\gmin{\gamma_{\rm min}}
\def\lmax{l_{\rm max}}
\def\nmax{n_{\rm max}}

\def\ApJ{ApJ}
\def\ApJL{ApJ Lett}
\def\ApJSS{ApJ Suppl. Ser.}
\def\AstJ{AJ}
\def\AA{Astron. Astrphys.}
\def\AJ{Astron. J.}
\def\ARAA{Ann. Rev. Astron. Astrphys.}
\def\MNRAS{MNRAS}
\def\Nature{Nature}
\def\NewA{NewA}
\def\PASJ{PASJ}

\def\d2dt#1{{ {\rm d}^2 #1 \over {\rm d}t^2 }}

\def\simless{\mathbin{\lower 3pt\hbox
   {$\rlap{\raise 5pt\hbox{$\char'074$}}\mathchar"7218$}}} %< or of order
\def\simgreat{\mathbin{\lower 3pt\hbox
   {$\rlap{\raise 5pt\hbox{$\char'076$}}\mathchar"7218$}}} %> or of order
\makeatletter

\long\def\@makecaption#1#2{%
  \small
  \vskip\abovecaptionskip
  \sbox\@tempboxa{#1: #2}%
  \ifdim \wd\@tempboxa >\hsize
    #1: #2\par
  \else
    \global \@minipagefalse
    \hb@xt@\hsize{\hfil\box\@tempboxa\hfil}%
  \fi
  \normalsize
  \vskip\belowcaptionskip}
\def\thebibliography#1{\section*{References\markboth
 {REFERENCES}{REFERENCES}}\list
  {}{\settowidth\labelwidth{0pt}\leftmargin\labelwidth
 \advance\leftmargin\labelsep
 \usecounter{enumi}\@bibsetup}
 \def\newblock{\hskip .11em plus .33em minus -.07em}
 \sloppy\clubpenalty4000\widowpenalty4000
 \sfcode`\.=1000\relax}

\def\@citex[#1]#2{\if@filesw\immediate\write\@auxout{\string\citation{#2}}\fi
  \def\@citea{}\@cite{\@for\@citeb:=#2\do
    {\@citea\def\@citea{\@citesep}\@ifundefined
       {b@\@citeb}{{\bf ?}\@warning
       {Citation `\@citeb' on page \thepage \space undefined}}%
{\csname b@\@citeb\endcsname}}}{#1}}

\def\@bibsetup{\itemindent=-\leftmargin \itemsep=0pt
 \parsep=0pt
 }

\def\@citesep{; }
\def\@cite#1#2{({#1\if@tempswa , #2\fi})}
\def\@biblabel#1{\hfill}

\makeatother
\begin{document}

\title{{\bf The dynamical evolution of massive black hole binaries - II.
Self-consistent N-body integrations.}}
\author{Gerald D.~Quinlan$^1$ and Lars Hernquist$^2$\\
$^1$Dept.\ of Physics and Astronomy, Rutgers University, PO Box 849,
Piscataway NJ 08855\\
(Present address: 1055 Bay St., \#1503, Toronto, Canada \ M5S 3A3)\\
$^2$Lick Observatory, University of California, Santa Cruz CA 95064}

\maketitle

\begin{abstract}
We use a hybrid N-body program to study the evolution of massive black hole
binaries in the centers of galaxies, mainly to understand the factors
affecting the binary eccentricity, the response of the galaxy to the binary
merger, and the effect of loss-cone depletion on the merger time.  The
scattering experiments from paper~I showed that the merger time is not
sensitive to the eccentricity growth unless a binary forms with at least a
moderate eccentricity.  We find here that the eccentricity can become large
under some conditions if a binary forms in a galaxy with a flat core or with
a radial bias in its velocity distribution, especially if the dynamical
friction is enhanced by resonances as suggested by Rauch and Tremaine
(1996).  But the necessary conditions all seem unlikely, and our prediction
from paper I remains unchanged: in most cases the eccentricity will start
and remain small.  As a binary hardens it ejects stars from the center of a
galaxy, which may explain why large elliptical galaxies have weaker density
cusps than smaller galaxies.  If so, the central velocity distributions in
those galaxies should have strong tangential anisotropies.  The wandering of
a binary from the center of a galaxy mitigates the problems associated with
loss-cone depletion and helps the binary merge.
\end{abstract}

%==============================================================================
\section{Introduction}
%==============================================================================

The formation of massive black hole (\BH) binaries follows from two widely
held assumptions: that galaxies merge and that many galaxies contain central
\BHs.  The evolution of these binaries is relevant to a number of problems in
extragalactic astronomy.  They have been proposed to explain the bends in
jets from active galaxies (Begelman et al., 1980; Roos, 1988; Roos et al.,
1993; Gaskell, 1996), and the difference between radio-loud and radio-quiet
active galaxies (Wilson and Colbert, 1995).  Multiple mergers of \BH\
binaries can lead to the buildup of massive central \BHs\ (Hut and
Rees, 1992), although if the binaries do not merge quickly the \BHs\
can be ejected through three-body interactions (Xu and Ostriker, 1994;
Valtonen, 1996), a process that has been proposed to explain the double
radio lobes of active galaxies (Saslaw et al., 1974; Valtonen et al.\ 1994).
The double nucleus of the nearby galaxy M31 (Lauer et al., 1993) could be a
\BH\ binary in the making, although Tremaine (1995) has proposed a plausible
model that requires only one \BH.  If binary mergers are frequent they will
be the most interesting source of gravitational radiation for the planned
space-based detector LISA (Haehnelt, 1994; Bender et al., 1995).  Lastly,
the evolution of \BH\ binaries may explain why the density profiles of
elliptical galaxies fall into two classes, with small galaxies having strong
density cusps and large galaxies having weaker cusps (Ebisuzaki et al.,
1991; Faber et al., 1996).

The relevance of \BH\ binaries to all of these problems depends on how long it
takes for dynamical friction (resulting from star-\BH\ interactions) and
gravitational radiation to cause the binaries to merge.
The merger time is difficult to compute because the hardening of the binary
changes the structure of the galaxy, which in turn changes the hardening
rate.
The back-of-the-envelope calculations of Begelman et al.\ (1980) have now
been supplemented with numerical results from three-body scattering
experiments (Roos, 1981; Mikkola and Valtonen, 1992) and full N-body 
experiments (e.g., Ebisuzaki et al., 1991; Makino, 1997), yet several big
uncertainties remain. 
The eccentricity evolution has been subject to much debate,
the most recent contribution being that of Rauch and Tremaine (1996) who
argue that resonant interactions between the \BHs\ and the stars can cause a
rapid growth in the eccentricity if the velocity distribution has a radial
anisotropy.  
This would cause gravitational radiation to become important early in the
evolution and would help the binaries merge.
Another uncertainty has been loss-cone depletion, the depletion of the
low-angular-momentum stars that can interact with the binary,
which lowers the hardening rate and lengthens the merger time.
Because of the uncertainties one can find papers in the recent literature
making widely different predictions for merger times in typical galaxies,
ranging from $\simless 10^8\,$yr (e.g., Makino and Ebisuzaki, 1994) to well
over a Hubble time (e.g., Rajagopal and Romani, 1995; Valtonen, 1996).

In paper I (Quinlan, 1996) we presented results from a new set of three-body
scattering experiments as a first step towards resolving the uncertainties.
We computed the hardening rate, the eccentricity growth rate, and the
rate at which a binary ejects stars, and derived formulas to fit the
dependence of the rates on the hardness and mass ratio of the binary.
One of the main results was that the eccentricity of a hard binary
grows only slowly as the binary hardens, and that the growth does not
change the merger time by much unless the binary starts with at least a
moderate eccentricity, say $e\simgreat0.5$ (this had been demonstrated
for equal-mass binaries by Mikkola and Valtonen, 1992).
But in deriving these results we made some simplifying assumptions that
may not be applicable to real galaxies: we assumed a homogeneous,
isotropic galaxy core with a Maxwellian velocity distribution, and we
ignored the change in the core caused by the binary.
We thus could not study the response of the galaxy and the depletion of
loss-cone orbits in a self-consistent manner, nor could we study the
dependence on the initial density profile or on anisotropies in the velocity
distribution.

To improve upon the scattering experiments we have developed a hybrid N-body
program that is well suited to studying the evolution of \BH\ binaries
in models for elliptical galaxies, including models with cusps and
anisotropies.
In this paper we describe the design and performance of the program and
then use it to study the factors affecting the eccentricity, the response of
the galaxy to the binary hardening, and the effect of loss-cone depletion on
the hardening rate. 
We pay special attention to the convergence of the results with the number
of particles and to some differences between the N-body results and the
predictions from paper~I.

We ignore some complications that may be important for real galaxies.
We consider only spherical or nearly spherical galaxies, and ignore 
disks or triaxial components of the potential.
A disk will alter the frictional force on the \BHs, possibly dragging them
into the disk plane, and a triaxial component will help the merger by
increasing the number of stars that can interact with the binary.
We also ignore gas, which can help the merger by accreting onto the \BHs\ or
by applying a frictional force to the \BHs.
We wish to understand the simple problem with purely stellar-dynamical
processes first before we add complications.
The neglected complications should not matter for binaries in gas-poor,
nearly spherical galaxies like M87.

%==============================================================================
\section{Computational methods}
%==============================================================================

%------------------------------------------------------------------------------
\subsection{The N-body program}
%------------------------------------------------------------------------------

N-body programs are usually designed to solve problems of one of two types:
collisional or collisionless.
In collisional problems the evolution depends in an essential way on close
encounters between the stars. These problems require accurate integrations
with the exact $1/r^2$ force law, and are usually solved with high-order
integrators.
In collisionless problems the evolution depends not on close encounters
between the stars but on the collective response of the system to
changes in the potential. These problems require a large number of particles
to suppress spurious relaxation, and are usually solved with a low-order
integrator (e.g., leapfrog) and a fast, approximate method for computing the
forces.
The \BH-binary problem does not fit neatly into either of these types.
The binary evolves because of close encounters between the \BHs\ and the
stars, which must be integrated accurately, but the galaxy evolves because
of the collective response of the stars to changes in the potential, and the
number of stars must be large, both to suppress relaxation and to satisfy
the requirement that the stars be much less massive than the \BHs.

It is not surprising then that both types of N-body programs have been used
for the problem.
Governato et al.\ (1994) used a tree program with the leapfrog integrator
and a large softening length to study the merger of King models containing
central \BHs, with 16,000 particles per galaxy.  Their program was
good for the galaxy merger and the early stages of the binary evolution, but
not for the late stages, which they did not study.
Mikkola and Valtonen (1992) used Aarseth's NBODY1 program to study the
evolution of a \BH\ binary at the center of a Plummer model with 10,000
particles. 
NBODY1 uses a fourth-order integrator with individual stepsizes, which makes
it better than a tree program for the late stages of the evolution,
but it computes the forces by direct summation, which is slow.
Mikkola and Valtonen assumed a fixed Plummer potential when computing the
forces between the stars, which made the integrations go faster but at the
price of ignoring the self-consistent response of the galaxy.
Self-consistent integrations of galaxy mergers with central \BHs\ using over
$10^5$ particles per galaxy were done with a direct-summation program by
Makino and Ebisuzaki (1996) and Makino (1997), but on a GRAPE-4
supercomputer, whose speed of 100--150~Gflops is beyond the reach of 
general-purpose computers.

Our hybrid N-body program works well for both the collisional and
collisionless aspects of the \BH-binary problem.
We call it SCFBDY because it combines parts from the SCF program of
Hernquist and Ostriker (1992) and the NBODY\ programs of Aarseth (1994).
We describe here the main features of the program and their advantages for
the \BH-binary problem; a more technical description is given in the appendix.

The program moves the \BHs\ and stars according to different equations
of motion (we use lower-case letters for the stars, upper-case for the \BHs,
and set $G=1$):
%---------------------------- old--------------------------------------------
%                             Problem: this will not fit in two-column format.
% \begin{eqnarray}
%    \d2dt{\vR_i} &=& - \sum_j { M_j (\vR_i-\vR_j) \over |\vR_i-\vR_j|^3 }
%                     - \sum_j { m_j (\vR_i-\vr_j) \over 
%             ( |\vR_i-\vr_j|^2 + \epsilon_i^2 )^{3/2} }      \label{eq-bh}\\
%    \d2dt{\vr_i} &=& - \sum_j { M_j (\vr_i-\vR_j) \over 
%               ( |\vr_i-\vR_j|^2 + \epsilon_j^2 )^{3/2} }
%                    - \sum_{n,l,m} A_{nlm}\nabla\Phi_{nlm}    \label{eq-st} 
% \end{eqnarray}
%---------------------------- old --------------------------------------------
\begin{equation}                                                \label{eq-bh}
   \ddot{\vR_i} = \sum_j { M_j (\vR_j-\vR_i) \over |\vR_j-\vR_i|^3 }
                     + \sum_k { m_k (\vr_k-\vR_i) \over 
             ( |\vr_k-\vR_i|^2 + \epsilon_i^2 )^{3/2} },
\end{equation}
\begin{eqnarray}                                                \label{eq-st}
   \ddot{\vr_k} = \sum_j { M_j (\vR_j-\vr_k) \over 
              ( |\vR_j-\vr_k|^2 + \epsilon_j^2 )^{3/2} }
                   - \sum_{n,l,m} A_{nlm}\nabla\Phi_{nlm}    
\end{eqnarray}
The \BHs\ interact with the other \BHs\ and with the stars through the
$1/r^2$ force (a softening parameter is added for the \BH-star interactions).
The stars interact with the \BHs\ through the $1/r^2$ force, but with the
other stars through an expansion of the potential with coefficients
$A_{nlm}$ that are updated with time self-consistently.   
By using the $\Phi_{nlm}$ basis functions of Hernquist and Ostriker (1992)
we can fit the potentials of most elliptical-galaxy models adequately with a
small number of coefficients---a dozen is often enough for spherical
galaxies.
Phinney and Villumsen (Phinney 1994) split the equations for the \BH-binary
problem in a manner similar to our equations (\ref{eq-bh}) and
(\ref{eq-st}), but used a different expansion for the potential and used the
leapfrog integrator to move the particles. 
We move the particles with individual stepsizes (sometimes differing by as
much as a factor of $10^6$) using one of Aarseth's (1994) fourth-order
integrators---the NBODY1 integrator for the stars, and NBODY2 or NBODY6
for the \BHs.
Our program thus combines the integration accuracy of a direct-summation
program with the speed of a SCF type of program.

The program does have some limitations.  It does not vectorize or
parallelize as well as the pure SCF program.   
It modifies two-body interactions between the stars by replacing the
potential by a smooth approximation, and hence cannot be used to study the
diffusion of stars into the loss cone by two-body relaxation.
(Programs that do not replace the potential by a smooth approximation commit
the opposite error for the \BH-binary problem: they give a diffusion rate
larger than would occur in a real galaxy, because they cannot use as many
stars as there are in a real galaxy.) 
And the program requires a central point about which the potential can be
expanded, at least in the simple version used here with only one expansion
center, which means that it cannot be used to study the galaxy mergers that 
lead to galaxies with more than one central \BH.
We ignore that stage of the evolution.

%------------------------------------------------------------------------------
\subsection{Initial conditions and units of measurement}
%------------------------------------------------------------------------------

We used three simple, isotropic galaxy models for our experiments in this
paper, as well as some anisotropic models that will be described later. 
The first of the simple models is the Plummer model, with a density 
\begin{equation}                                                \label{eq-plum}
   \rho(r) = {3M \over 4\pi d^3} \left( 1 + {r^2\over d^2}\right)^{-5/2}.
\end{equation}
The Plummer model is easy to set up and interpret, but is a poor model for
real galaxies: they are much more concentrated, many with densities that
continue rising to the innermost observable radius (Crane et al., 1993;
Ferrarese et al., 1994; Forbes et al., 1995; Lauer et al., 1995). 
We therefore used two models with density cusps from the family considered
by Dehnen (1993) and Tremaine et al.~(1994):
\begin{equation}                                               \label{eq-gamma}
   \rho(r) = {3-\gamma \over 4\pi} { Md\over r^\gamma (r+d)^{4-\gamma} }.
\end{equation}
We used the models with $\gamma=1$ and $\gamma=2$, which are the models of
Hernquist (1990) and Jaffe (1983).
The ``$\gamma$~models'' fit density profiles of elliptical galaxies
reasonably well if the length scale $d$ is chosen to be close to the
effective radius.
The inner parts of elliptical galaxies can be fit better by Zhao's (1996)
formula, $\rho(r) \sim r^{-\gamma} (r^{1/\alpha}+d)^{(\beta-\gamma)\alpha}$,
with $\beta$ typically in the range 2--3 (the length scale $d$, or ``break
radius'', is much smaller than the effective radius in these fits), which is
similar to the fitting formula used by Byun et al.\ (1996).
The difference between Zhao's models and the $\gamma$ models is unimportant
for our work because we are mainly interested in the region $r< d$ where the
models are the same. 
We truncate the models at $r=300d$.

To increase the statistical resolution near the center, and to help satisfy
the requirement that the stars be much less massive than the \BHs, we give
the stars a mass spectrum, with the masses decreasing towards the center as 
\begin{equation}                                              \label{eq-lambda}
   m\sim r_p^\lambda, 
\end{equation}
with $r_p$ the initial pericenter of the star and $\lambda$ an exponent,
usually in the range 0.5--1.0 (Sigurdsson et al., 1995).  In a Jaffe model
with $\lambda=0.75$, for example, about 40\% of the stars have $r/d<0.1$ and
20\% have $r/d<0.01$, whereas the corresponding percentages for an
equal-mass model are 10\% and 1\%.  The mass spectrum does not lead to mass
segregation, because the potential is smooth with our program and the
segregation time is long.  We impose minimum and maximum masses, typically
differing by a factor of 1000, so that the masses do not become arbitrarily
small or large at small and large radii.

The \BHs\ are treated as point masses obeying Newtonian gravity.
The detected massive \BHs\ in real galaxies have a mean mass of 0.002--0.003
(in units of the bulge mass), although there is a scatter of at least one
order of magnitude about this mean (Kormendy and Richstone, 1995).
It is difficult to use small \BHs\ in the experiments because the
dynamical-friction time for those is long and because we need the stars
to be much less massive than the \BHs.
We therefore use \BHs\ somewhat larger than occur in real galaxies,
$\mbh/M\simeq 0.01$, and vary their masses to determine how the results
scale with the mass. 
The scaling is not important for the eccentricity results, which are not
sensitive to the masses (except for the results on resonant friction), but
it is important for the response of a galaxy to the binary evolution. 
Our masses are not that unrealistic given that our galaxy models have less
mass at large radii than a real galaxy would have (if we interpret $r=d$ as
the break radius).

The initial condition for our \BHs\ are of two types.
In the first we start the \BHs\ symmetrically about the origin in one of the
galaxy models described above, at $r\simeq d$.
We assume that the \BHs\ arrived there after a galaxy merger, or after being
dragged in from the halo of the galaxy.
After a merger the \BHs\ would probably be surrounded by clusters of
bound stars, but we ignore those (our potential expansion has trouble
resolving small clumps of stars away from the center).
In the second we start a galaxy with one large \BH\ at the center,
surrounded by a cluster of bound stars, and allow a smaller \BH\ with no
cluster of bound stars to sink in from $r\simeq d$.
The second type of initial condition is more realistic than the first, but
the first is adequate for answering some questions and is easier to use than
the second, because in the second the central \BH\ and the stars surrounding
it require small stepsizes right from the start of the integration.

We present the N-body results in a system of units in which $G=M=1$, with
$M$ the mass of the galaxy without the \BHs.
We usually choose the length scale so that $E=-1/4$, with $E$ the energy of
the initial galaxy without the the \BHs. 
The length scales $d$ for the Plummer and $\gamma$ models are then $3\pi/16$
and $1/(5-2\gamma)$, and the half-mass radius and the density and dynamical
time at that radius are all of order unity.
We specify the softening for the \BH-star interactions by either the
softening length $\soft$ or the softening velocity  
\begin{equation}                                               \label{eq-vsoft}
   \vsoft = \sqrt{G\mbh/\soft}.
\end{equation}
As in paper~I, we define the binary orbital velocity $\Vbin$ by
\begin{equation}                                               \label{eq-vbin}
   \Vbin = \sqrt{G\M12/a},
\end{equation}
with $\M12=m_1+m_2$.

%==============================================================================
\section{A sample integration}                               \label{sec-sample}
%==============================================================================

We shall describe one N-body experiment in detail to explain our
assumptions, integration procedure, and method of data analysis, and to
identify some of the questions to be studied later.  We also use this
experiment to test the sensitivity of the results to the integration
parameters and the number of particles.

The problem that we consider is the hardening of an equal-mass binary at the
center of a Jaffe model.  The galaxy was represented by $10^5$ particles
with a mass exponent $\lambda=0.75$.  The \BHs\ had masses $m_1=m_2=0.01$, and
were started on nearly circular orbits at $(\vr,\vv)$ and $(-\vr,-\vv)$,
with $r=0.5$.  We integrated the orbits for 80 time units with an accuracy
parameter $\eta=0.005$, a softening velocity $\vsoft=16$, and a spherical
potential expansion (with $\nmax=18$, $\lmax=0$) whose coefficients were
updated every $\Delta t=0.005$; the integration took five days of cpu time
on a DEC 250-4/266 AlphaStation.  The program regularized the orbits at time
$t=17.85$.  The total energy was conserved to about one part in 100.

\begin{figure*}[tb]
\centerline{\psfig{figure=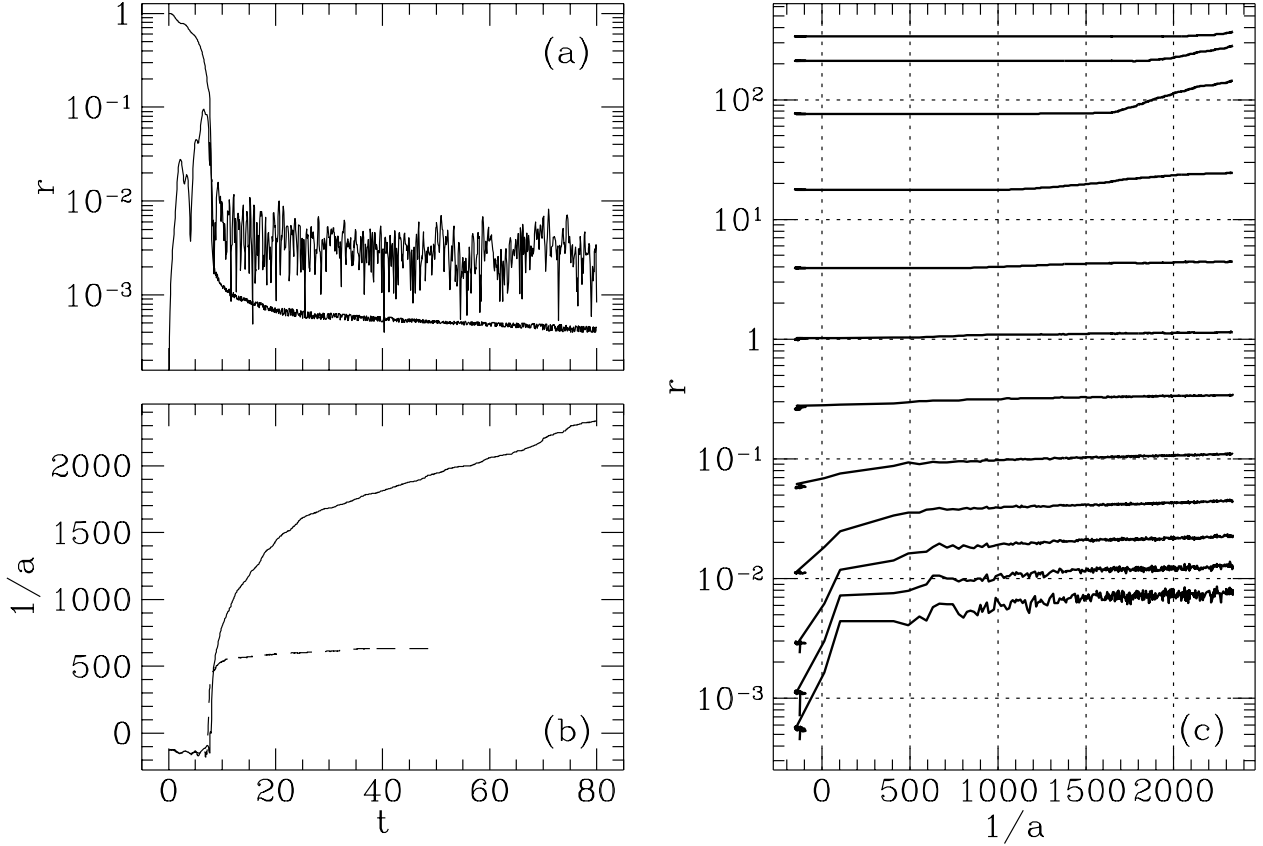,width=1.02\textwidth,clip=}}
\caption[Binary evolution in a Jaffe model.]
{Binary evolution in a Jaffe model. Two equal-mass \BHs\ ($m_1=m_2=0.01$) were
started on nearly circular orbits at ($\vec r,\vec v$) and ($-\vec r,-\vec
v$), with $r=0.5$.  The panels show (a) the separation between the \BHs\ (the
lower line towards the end) and between their center of mass and the origin,
(b) the semimajor axis, and (c) the radii containing fixed percentages of
the galaxy mass: 0.008 (bottom line), 0.04, 0.2, 1, 5, 20, 50, 80, 95, 99,
99.8, and 99.96 (top line). The dashed line in panel~(b) is from an
integration in which the center of mass of the binary was fixed at the
origin.}
\label{fig-evjaf}
\end{figure*}

The integration results are shown in Figure~\ref{fig-evjaf}.
%++++++++++++++++++++++++++++++++++++++++++++++++++++++++++++++++++++++++++++
%++++++++++++++++++++++++++++++++++++++++++++++++++++++++++++++++++++++++++++
The semimajor axis $a$ and eccentricity $e$ were computed from
the \BH\ coordinates and velocities by the usual formulas for a Kepler orbit,
ignoring the interaction between the \BHs\ and the stars.
The eccentricity remained small ($e\simless 0.1$) and is not shown in the
figure. 
The semimajor axis is shown in panel~(b).
The passage of $1/a$ through zero marks the time when the \BHs\ become bound,
$t=6.15$, at which the separation between the \BHs\ is $r=0.063$.
A \BH\ binary becomes hard once its semimajor axis shrinks to $\ah =
Gm_2/4\sigma^2$ (Quinlan, 1996), with $\sigma$ the one-dimensional dispersion
near the center but outside the Keplerian rise around the \BHs. 
For the Jaffe model, $1/\ah=200$ ($\sigma^2 = 1/2$).
The hardening slows down noticeably after $1/a$ reaches about 500, because
the binary has by then ejected many stars and has reduced the central density.
But the hardening does not stop, because the binary does not remain at the
center: it wanders around and is thus able to interact with new stars.
This is apparent from the plot of the center-of-mass radius in panel~(a).

To check the importance of wandering we repeated the integration with a
constraint force applied to keep the center of mass of the binary fixed at
the origin. We did this by moving the \BHs\ with the same stepsizes and by
replacing the forces $\vec F_1$ and $\vec F_2$ on the \BHs\ at each step by 
$\vec F_1 - \vec \Fcm$ and $\vec F_2 - \vec \Fcm$, with 
$\vec \Fcm=(m_1\vec F_1 + m_2\vec F_2)/\M12$. 
With this constraint force the hardening stopped after $1/a$ reached about
650---see the dashed line in panel~(b).
A series of integrations like this with different binary masses showed that
the hardening stops when the mass in stars that can approach within $a$ of
the center drops to a small fraction of the binary mass. 
For the Jaffe model the final, ``loss-cone'' separation $\alc$ therefore
varies linearly with the binary mass, $\alc\sim\M12$.  
This differs from the prediction of Begelman et al.~(1980), but is similar
to the predictions made by Roos~(1981).
When the binary is free to wander the hardening does not stop; we shall say
more about this later.

We stopped the integration at $t=80$ because it was becoming slow, owing to
the small stepsizes of the \BHs.
In a real galaxy gravitational radiation would at some point cause the \BHs\
to merge.
The time at which that would happen in our experiment depends on the
physical mass and length scales, which we have not specified.
Our neglect of gravitational radiation is
justified during most of the evolution because the importance of radiation
turns on suddenly---the timescales for hardening by three-body
scattering and by radiation vary as $\sim 1/a$ and $\sim a^4$.
Our goal is to integrate the Newtonian equations far enough that we can
predict how the evolution would continue until the point at which radiation
becomes important.
For the applications that we are interested in---\BHs\ of mass
$10^8$--$10^9\,M_\odot$ in a large galaxy like M87, or \BHs\ of mass
$10^5$--$10^6\,M_\odot$ in a smaller, denser galaxy like M32---radiation
becomes important after the semimajor axis shrinks by a factor of 10--50
beyond $a=\ah$ (Quinlan, 1996). 
In our integration the semimajor axis shrank by a factor of~12 beyond
$a=\ah$, which might have been enough.

Panel~(c) shows how the galaxy expands as the stars absorb energy from the
\BHs. 
This starts when the \BHs\ first sink into the center, before they become
bound, but most of it occurs after the binary forms and begins hardening. 
The expansion appears to slow down near the end, but that is mainly because
the figure uses a linear scale for $1/a$ and a logarithmic scale for $r$.
Near the end the outer parts of the galaxy begin expanding too, because of
the stars that the binary has ejected.

To test the dependence of the results on the integration parameters we 
repeated the integration with different choices of $\eta$, $\vsoft$, $\Delta
t$, $\nmax$, and $\lmax$. 
None of the changes had a significant effect on the binary evolution unless
a parameter was changed to an unreasonable value, such as a softening
velocity lower than the orbital velocity of the binary at the end of the
integration ($\Vbin= 6.8$), or a value of $\eta$ or
$\Delta t$ so large that energy conservation was grossly violated.
Even adding non-spherical terms to the potential by changing $\lmax$ from
0 to 2 hardly affected the binary, because the galaxy remained nearly
spherical throughout the integration.
The parameters that we used thus seem adequate for our purposes.

%++++++++++++++++++++++++++++++++++++++++++++++++++++++++++++++++++++++++++++
\begin{figure*}[htb]
\centerline{\psfig{figure=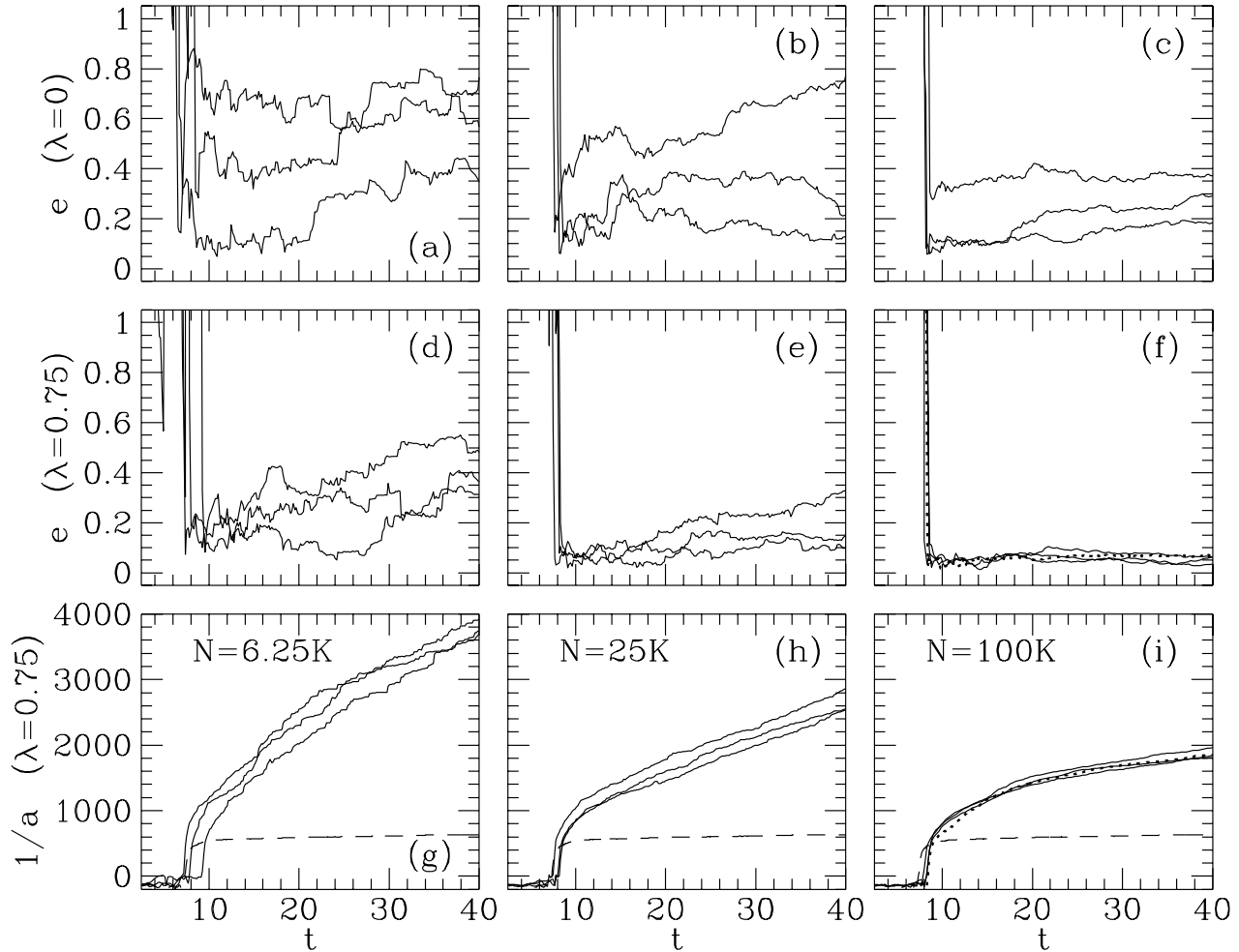,width=1.02\textwidth,clip=}}
\caption[Dependence of the evolution on the number of particles ($N$)]
{Dependence of the binary evolution on the number of particles ($N$), for
the experiment shown in Fig.~\ref{fig-evjaf}. The left, middle, and right
columns show the eccentricity and semimajor axis computed with
$N/10^3=6.25$, 25, and 100. The integrations in the top row used equal-mass
particles ($\lambda=0$); those in the bottom two rows used a mass spectrum
($\lambda=0.75$).  For each combination of $\lambda$ and $N$ the integration
was done three times with three different orbital planes for the initial
orbits of the \BHs.  The dotted lines in panels (f) and (i) are from an
integration with $N=2\times10^5$ and $\lambda=0.75$. The dashed lines in the
bottom row are from panel~(b) of Fig.~\ref{fig-evjaf}.}
\label{fig-eavsn}
\end{figure*}
%++++++++++++++++++++++++++++++++++++++++++++++++++++++++++++++++++++++++++++

To test the dependence on the number of particles  we compared results
from integrations using 6250, 25000, and $10^5$ particles, using both
equal-mass particles ($\lambda=0$) and particles with a mass spectrum
($\lambda=0.75$).
For each $(\lambda,N)$ combination we performed three integrations, with the
orbits of the \BHs\ started in the XY, YZ, and ZX planes; the difference
between them is a measure of the noise level.
The results are summarized in Figure~\ref{fig-eavsn}.

The eccentricity is noisy when $N$ is low, especially with the models with
equal-mass particles, because the eccentricity of a hard binary gets
perturbed only by a small fraction of the stars---those that have close
encounters with the binary.
In the models with a mass spectrum that fraction is larger and the
eccentricity is less noisy.
The convergence of the eccentricities towards $e\simeq0.05$ as $N$ rises in
panels (d), (e), and (f) suggests that it is the correct answer.

The semimajor axis is not as noisy as the eccentricity.
Even with only 6250 particles the integrations starting from the three
orbital planes give nearly the same hardening rate.
But there is a systematic decrease in the rate as $N$ rises from 6250 to
$10^5$. 
A similar decrease was observed by Makino (1997), who proposed two
explanations: two-body relaxation and the wandering of the binary from the
center of the galaxy. 
Since our program suppresses two-body relaxation, the decrease most likely
results from interactions between the stars and the center of mass of the
binary. 
As $N$ rises the perturbations that the binary experiences become less noisy
and the binary wanders less from the center.
It therefore reduces the density near the center more and its hardening
slows down.
The wandering amplitude cannot be made arbitrarily small by raising $N$,
as it could if the binary were a single particle, because the restoring
force that keeps the binary at the center grows weaker as the central
density goes down. 
There is therefore an $N$ value above which the hardening stops decreasing
with $N$, about $10^5$ for the experiment considered here: panel~(i) shows
that the hardening rate with $N=2\times10^5$ is the same as with $N=10^5$. 
For smaller \BHs\ the wandering continues growing to larger $N$ values.
The mass spectrum affects the wandering less than it affects the
eccentricity, because the center of mass gets perturbed by all the stars
near the center, not just by the stars having close encounters with the
binary.  The mean hardening rates found with equal-mass particles differed
only slightly (they were larger) from those found with a mass spectrum.

This sample integration has thus shown that our program can integrate
binaries accurately in large N-body systems without great expense, but also
that the results have to be extrapolated with care before they can be
applied to real galaxies.
The value of $N$ can easily be the largest source of error if it is small.  
For the eccentricity it is possible with the help of a mass spectrum to
choose $N$ large enough to suppress the noise---the value required is
inversely proportional to the masses of the \BHs.
For the semimajor axis this is more difficult, because the mass spectrum
does not help as much, but it is still possible if the \BHs\ are large.
The main uncertainty that the value of $N$ introduces is in the hardening
rate; it does not affect the response of the galaxy (except right at
the center) if that is plotted versus the binary semimajor axis as it is in
panel~(c) of Figure~\ref{fig-evjaf}.

%==============================================================================
\section{The effect of wandering on the hardening and ejection rates}
\label{sec-wand}
%==============================================================================

Our N-body program can be used both to test the predictions from our 3-body
scattering experiments and to study dynamical processes that could not be
explored with those experiments.  In this section we test the predictions
for the hardening and mass-ejection rates.  There are several reasons for
suspecting that these rates may differ in the N-body experiments.  The most
obvious is that the stellar distribution evolves along with the binary in
the N-body experiments, whereas in the scattering experiments it is assumed
to be fixed.  When the low-angular-momentum parts of the distribution get
depleted the hardening slows down.  But differences can occur even in the
absence of loss-cone depletion, because in the N-body experiments the stars
are influenced by forces from both the binary and the galaxy, not just from
the binary as in the scattering experiments, and because in the N-body
experiments the binary does not remain fixed in space.  We argue here that
the wandering of the binary increases the ejection rate by about a factor of
two.

\begin{figure*}[tb]
\centerline{\psfig{figure=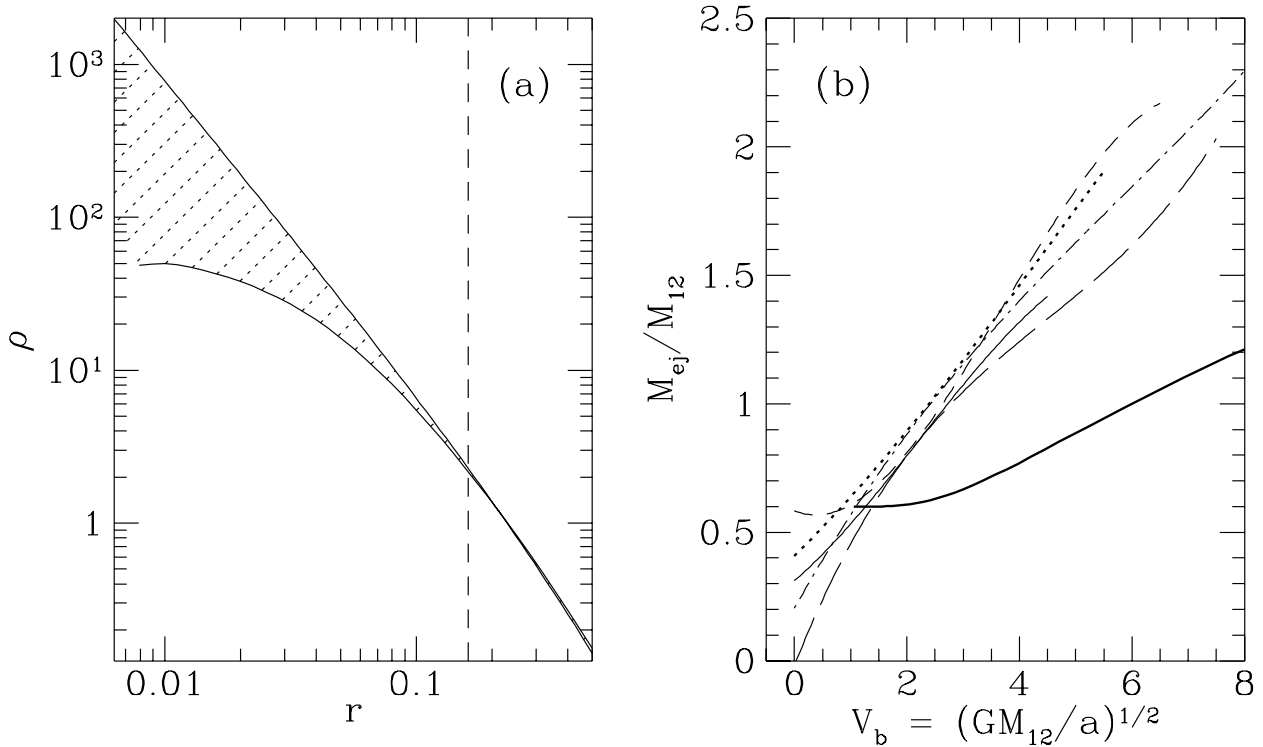,width=1.02\textwidth,clip=}}
\caption[Mass ejection.]
{Ejected mass during integrations of equal-mass binaries in a Jaffe model.
(a) Stellar density at the start (top line) and end (bottom line) of the
integration of Fig.~\ref{fig-evjaf}. The dashed line is drawn at $r=r_*$;
the shaded region indicates the ejected mass.  (b) Ejected mass versus 
binary orbital velocity from integrations like that of Fig.~\ref{fig-evjaf}
but with $m_1=m_2=$ 0.04 (solid line), 0.02 (dotted), 0.01 (short dashed),
0.005 (long dashed), and 0.0025 (dashed dotted); the heavy solid line is the
prediction from the scattering experiments of paper I.}
\label{fig-mejj}
\end{figure*}

The evidence for this is shown in Figure~\ref{fig-mejj}.
%++++++++++++++++++++++++++++++++++++++++++++++++++++++++++++++++++++++++++++
%++++++++++++++++++++++++++++++++++++++++++++++++++++++++++++++++++++++++++++
In panel~(b) we plot the ejected mass as a function of the binary orbital
velocity during integrations of binaries in a Jaffe model (with $N=10^5$,
and with the binaries started as was the binary in Fig.~\ref{fig-evjaf}).
We computed the ejected mass at time $t$ from $\Mej(t) = M(r_*,0) -
M(r_*,t)$, with $r_*$ the radius at which $\rho(r_*,t)/\rho(r_*,0) = 0.95$.
The five lines in panel~(b) are about the same within the uncertainties; we
take the dashed-dotted line as the mean N-body result (integrations with
other $N$ values gave similar results).  We also plot the prediction from
our 3-body scattering experiments, computed from
\begin{equation}
   \Mej(a)/\M12 = C + \int^\infty_a\!{da\over a}\, J(a),
\end{equation}
where $C$, which we set to 0.6, is to allow for the mass that gets removed
when the \BHs\ first sink into the center, and $J$ is the 3-body ejection rate
(in what follows we speak of $d\Mej/d\Vbin$ loosely as though it were the
ejection rate).  In computing the 3-body prediction we assumed a Maxwellian
velocity distribution with $\sigma^2=1/2$, and, as we did in paper~I, we
counted stars as ejected if they get expelled from the binary with a
velocity larger than $\vej=\sqrt{12}\sigma$.  Although there is some
uncertainty in the appropriate values for $C$, $\sigma$, and $\vej$, it is
too small to explain the factor-of-two difference in the slopes of the
N-body and 3-body ejection curves near the end of the integration. Similar
differences were found from integrations with galaxy models with other
$\gamma$ values.

A simple thought experiment suggests that the wandering of the binary can
account for a large part of the difference.  Suppose the binary moves so
fast that incoming stars can interact with it only once before it moves out
of their reach.  For some stars this will not matter, but for others, which
would experience multiple interactions before getting expelled if the binary
were not wandering, it will.  The cross sections in Figure~4 of paper~I show
that the hardening rate is sensitive to these multiple scattering events.
Additional calculations from those cross sections show that the hardening
rate for a hard, equal-mass binary gets reduced by a factor of 7.5 when
stars are allowed to interact with the binary only once, whereas $d\Mej/dt$
gets reduced by at most a factor of two.  This thought experiment is
admittedly unrealistic, because in reality the binary will not be moving as
fast as we have assumed, but the qualitative conclusion drawn from it should
be correct: wandering will decrease the hardening rate more than $d\Mej/dt$,
and hence will increase the ejection rate when it is expressed as
$d\Mej/d\Vbin$ or $d\Mej/d\ln(1/a)$.

To test this conclusion we compared the ejection rates from N-body
integrations with the binaries free to wander with those from integrations
with the center of mass of the binaries fixed at the origin (as was
described in Sec.~\ref{sec-sample}).  We added a small triaxial perturbation
to the potential (the same for both sets of integrations) to allow more
stars to reach the center, for otherwise in the integrations with the center
of mass fixed the hardening would not have continued far enough for us to
make a fair comparison.  The ejection rate was close to the 3-body
prediction when the center of mass was fixed, but not when it was free; the
difference was again about a factor of two.

If our explanation is correct, the hardening rate in the N-body experiments
should be lower than was predicted from the scattering experiments.  To test
this we performed integrations of equal-mass binaries in a Plummer model,
starting the binaries as did Mikkola and Valtonen (1992).  The Plummer model
is better than the Jaffe model for this test because its large, flat core
simplifies the 3-body prediction.  Mikkola and Valtonen said that their
N-body hardening rate agreed with the 3-body prediction, but we found a
lower hardening rate (we checked the dependence on all our integration
parameters); their rate might have been too large by chance, since the
results are noisy with only $10^4$ particles, the number that they used.  To
make a better comparison of the 3-body and N-body rates we performed a
series of integrations like that of Mikkola and Valtonen but increasing the
particle number and decreasing the \BH\ masses by a factor of two each time,
i.e.\ using $N/10^4=1$, 2, 4, 8, 16, together with $m_1=m_2=1/100$, 1/200,
1/400, 1/800, 1/1600.  The importance of loss-cone depletion goes down as
$m_1$ is lowered with $Nm_1$ held fixed, because the mass of the binary goes
down but the wandering amplitude does not.  We found the N-body hardening
rate to be about 0.5--0.7 as large as the 3-body prediction.

This change to the 3-body hardening rate is not important for estimates of
merger times, given the much larger change that results from loss-cone
depletion (we describe later how wandering changes the hardening rate when
loss-cone depletion occurs).  But the change to the ejection rate is
important, because it means that before merging the binaries will eject
about twice as much mass as we predicted in paper~I. A question still to be
answered is whether this is true for binaries with unequal masses.

%==============================================================================
\section{The factors affecting the eccentricity evolution}      \label{sec-ecc}
%==============================================================================

We now study the factors affecting the eccentricity of a binary during the
early stages of the evolution.  We first consider binaries with equal or
nearly-equal masses, whose early evolution can be predicted by the usual
dynamical-friction formula, and then binaries with a large mass ratio, whose
eccentricity may be influenced by resonant interactions with the stars.

%------------------------------------------------------------------------------
\subsection{Non-resonant friction}
%------------------------------------------------------------------------------

Chandrasekhar's dynamical-friction formula can be used in most circumstances
to predict the loss rates for energy and angular momentum as a \BH\ binary
forms at the center of a galaxy.
The eccentricity is influenced by two competing factors.
The velocity dependence of the friction formula increases the frictional
force at apocenter relative to that at pericenter if the \BHs\ are moving fast,
and therefore favors an increase in the eccentricity (Fukushige et al., 1992). 
The density gradient in the galaxy has the opposite effect if the density
is higher at pericenter than at apocenter (Polnarev and Rees, 1994).
Anisotropy is a further complication.
Casertano et al.\ (1987) integrated the orbits of single
massive particles in isotropic and radially anisotropic galaxy models using
both Chandrasekhar's formula and N-body methods, and found that anisotropy
can counteract the density gradient and allow orbits to reach the center
without their eccentricity decreasing.
The eccentricity did not increase, however, unless the anisotropy was strong
enough to make the galaxy susceptible to the radial-orbit instability.

Since Chandrasekhar's formula is applicable only to the early stages of the
binary evolution, the predictions that it makes for the eccentricity need to
be checked by N-body experiments. 
Three previous studies have done this, all using a Plummer model to
represent the galaxy. 
Makino et al.\ (1993) did experiments with 16,000 particles to study the
formation and evolution of equal-mass binaries with \BH\ masses ranging from
0.01 to 0.16.
Their binaries sometimes formed with large eccentricities, but those
experiments started with the \BHs\ approaching the center on nearly radial
orbits. 
Phinney and Villumsen (Phinney, 1994) did a similar experiment but started
the \BHs\ with a larger angular momentum (with $N=4\times10^4$, $m_1=0.03$,
$m_2=0.02$) and found that the eccentricity remained small. 
Mikkola and Valtonen (1992) also started the \BHs\ with a large angular
momentum and found a small initial eccentricity, although their experiment
started with the binary nearly formed, and the noise level was high enough
(with $N=10^4$, $m_1=m_2=0.01$) that the eccentricity bounced around in a
random manner once the binary began to harden.
The results from these studies were thus consistent with the predictions
from Chandrasekhar's formula, but were noisy and did not explore the effect
of a density cusp or a radial anisotropy.

%++++++++++++++++++++++++++++++++++++++++++++++++++++++++++++++++++++++++++++
\begin{figure*}[tb]
\centerline{\psfig{figure=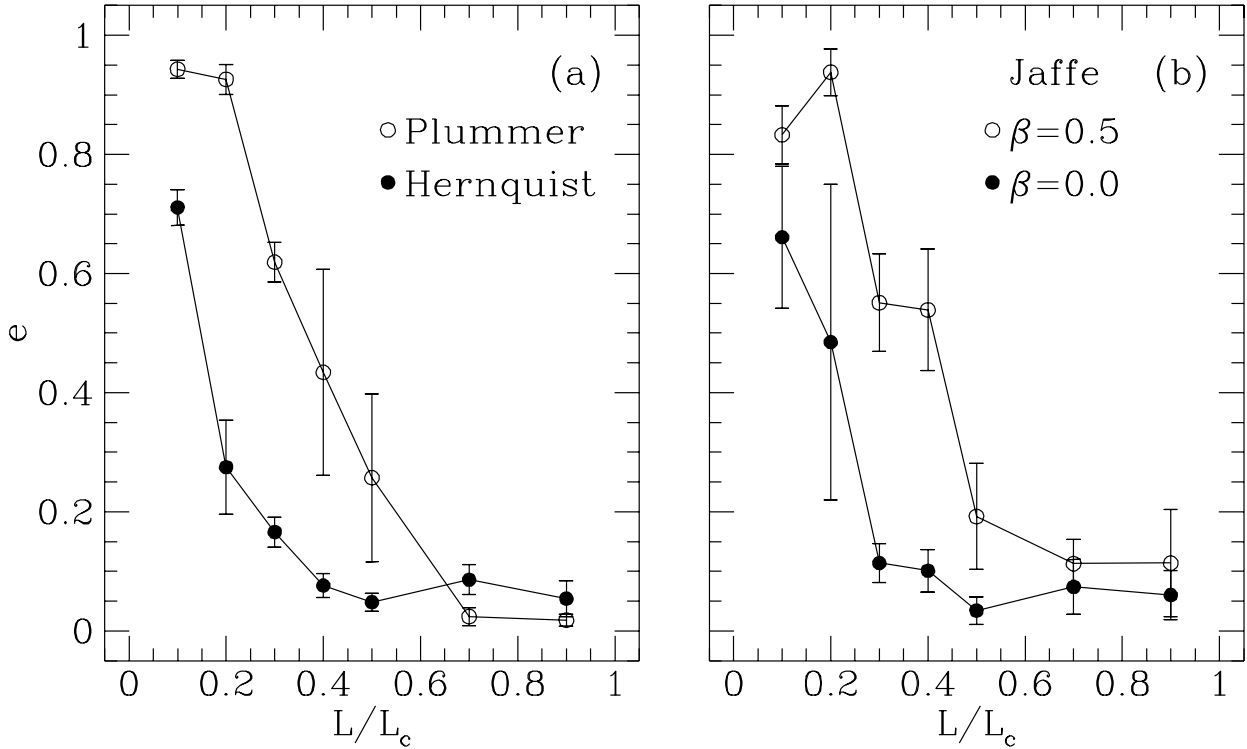,width=1.02\textwidth,clip=}}
\caption[Eccentricity versus angular momentum.]{
Eccentricity of an equal-mass binary when it first becomes hard at the
center of a galaxy.  The abscissa is the initial angular momentum of the \BHs\
(at $r=0.5$) in units of the angular momentum of a circular orbit of the
same energy.  The points and error bars show the mean eccentricity and the
average deviation from the mean when the semimajor axis reaches $a=0.004$,
computed from 3--5 experiments with different orbital planes for the initial
orbits.  The isotropic ($\beta=0$) Plummer and Hernquist models used
$N=10^5$ and $\lambda=1$; the Jaffe models used $N=2\times10^5$ and
$\lambda=0.5$.}
\label{fig-evsj}
\end{figure*}
%++++++++++++++++++++++++++++++++++++++++++++++++++++++++++++++++++++++++++++

To improve upon this we performed experiments like those of Makino et al.\
(1993) but using a variety of galaxy models and initial orbits, and
using a large number of particles to suppress the noise.
For the galaxy models we used those of Plummer, Hernquist, and Jaffe,
including for the Jaffe model both the isotropic model and a model with a
constant radial anisotropy of $\beta= 1-\sigma_t^2/\sigma_r^2=0.5$
(Cuddeford, 1991). 
This $\beta$ value is small enough to suppress the radial-orbit instability
(Merritt and Aguilar, 1985), which would be absent from our calculations in
any case because we include only spherical basis functions in the expansion
of the potential.
The \BHs\ had masses $m_1=m_2=0.01$ and were started symmetrically about the
origin at $(\vr,\vv)$ and $(-\vr,-\vv)$, with $r=0.5$ and with angular
momenta equal to $L/L_c$ times the angular momentum of a circular orbit of
the same energy.
The quantity that we measured is the eccentricity when the binary first
becomes hard, i.e.\ when it reaches $\ah=Gm_2/4\sigma^2$.

The results in panel~(a) of Figure~\ref{fig-evsj} show that it is 
easier for a binary to form with a large eccentricity in the Plummer model
than in the Hernquist model, in agreement with the prediction of Polnarev
and Rees~(1994) that a density cusp lowers the eccentricity. 
The isotropic Jaffe model is similar to the Hernquist model this respect
(the Jaffe-model results are noisier than the Hernquist-model results at
small $L/L_c$, possibly because of the different velocity distributions in
the two models).
For the anisotropic Jaffe model the eccentricities are larger at small values
of $L/L_c$ but are still small if $L/L_c\simgreat 0.6$, in agreement
with the prediction of Casertano et al.~(1987) that a radial anisotropy can
allow an eccentric orbit to remain eccentric but cannot cause the
eccentricity to grow.

If the incoming \BH\ orbits are distributed isotropically, the probability of
an orbit of a given energy having an angular momentum less than $L$ is
$(L/L_c)^2$.
The probability of a binary forming with a large eccentricity is therefore
small.
The formation of a \BH\ binary after the merger of two galaxies will be more
complicated than in the simple experiments considered here: the \BHs\ 
will have clusters of bound stars, and the surrounding galaxies will be
greatly disturbed.
Barnes (1992) simulated the merger of disk galaxies (without \BHs) and found
that the final collision of the two bulges was often remarkably head-on,
suggesting that small values of $L/L_c$ might not be that improbable.
But, as Barnes admits, the bulges in his galaxy models had unreasonably
large core radii.  
If the bulges had been more concentrated their final collision might have
had a larger $L/L_c$, for the same reason that the \BHs\ in our calculations
approach the center with a larger $L/L_c$ when the galaxy has a density cusp.
Thus, except for unusual cases in which a galaxy has a large, flat core or a
strong radial anisotropy, we suspect that most mergers of galaxies with
central \BHs\ will lead to binaries with small eccentricities.

%------------------------------------------------------------------------------
\subsection{Resonant friction}
%------------------------------------------------------------------------------

There is one circumstance in which dynamical friction can cause the
eccentricity of a \BH\ binary to increase even in the presence of a density
cusp. 
Rauch and Tremaine (1996) described this in their discussion of resonant
relaxation, and called the frictional force ``resonant friction'' because it
results from resonant interactions between the stars and the smaller of the
two \BHs.
Consider a \BH\ of mass $m_2$ orbiting a larger \BH\ of mass $m_1$, with $m_1$
surrounded by a cluster of bound stars of total mass $M_\star$, and assume
that the stars have an anisotropic velocity distribution ($df/dL\neq 0$).
If $m_1\gg \max(m_2,M_\star)$, the stars move in a potential that is nearly
Keplerian and have orbits that are nearly closed.
The frictional force on $m_2$ then cannot be described by the usual
dynamical-friction formula.
Rauch and Tremaine show that resonant interactions between $m_2$ and the
stars increase $|dL/dt|$ by a factor of order $m_1 / \max(m_2,M_\star)$ but
have no effect on $|dE/dt|$, and thus cause the eccentricity of the binary
to change much faster than the semimajor axis; the sign is such that the
eccentricity increases if $df/dL<0$ (i.e.\ if the distribution is biased in
favor of radial orbits) and decreases if $df/dL>0$.

\begin{figure*}[tb]
\centerline{\psfig{figure=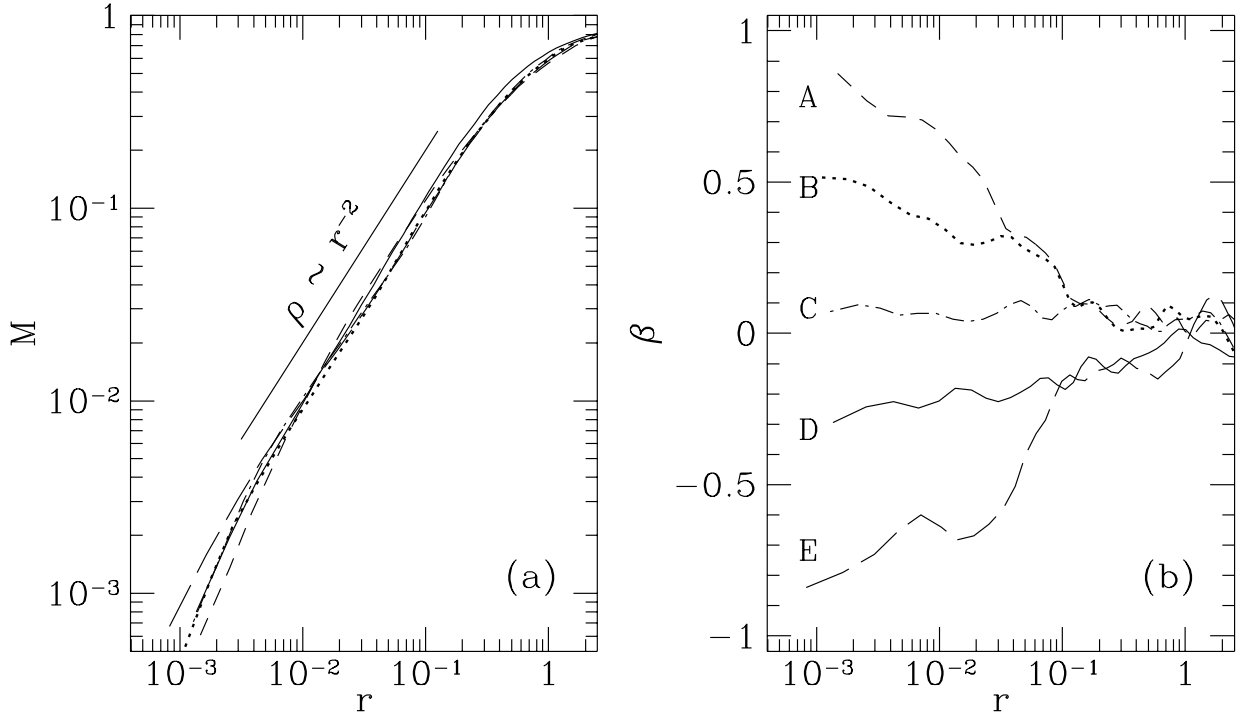,width=\the\hsize,clip=}}
\caption[Models for experiments with resonant friction.]{Anisotropy
$\beta = 1 - \sigma_t^2/\sigma_r^2$ and enclosed stellar mass $M$ for the
inner parts of five models used to study resonant friction. Each model has
$N=50$K particles with a mass exponent $\lambda=0.75$, and a central \BH\ of
mass $m_1=0.1$ with a \BH-star softening length of 0.00025.}
\label{fig-rfrmod}
\end{figure*}

To determine the strength of the anisotropy required for resonant friction
to be effective, we made five galaxy models having the same mass distribution
but different anisotropies, each model containing a central \BH\ of mass
$m_1=0.1$ (see Fig.~\ref{fig-rfrmod}).
We chose a large value for $m_1$ so that we could choose $m_2\ll m_1$
and still have $m_2$ much more massive than the stars; one can think of our
galaxy models in these experiments as representing the inner parts of real
galaxies.
Model~D was made by growing $m_1$ slowly at the center of a
$\gamma$~model with $\gamma=0$, which should lead to a final model with a
cusp $\rho\sim r^{-2}$ and a central anisotropy $\beta\simeq -0.2$
(Quinlan et al., 1995); the central $\beta$ value in the
figure is slightly lower than the theoretical prediction because we removed
some of the stars on radial orbits close to the \BH.
The other models were made by varying the cusp slope $\gamma$ of the
initial model and the growth time of the \BH\ (short growth times lead to
radial anisotropies, long growth times to tangential anisotropies), and
sometimes by removing stars on orbits with low or high angular momenta after
the \BH\ growth was finished to get the desired mass and anisotropy profiles.
The models were integrated for a while to check that they had reached
equilibrium.
Resonant friction should be important when a small mass $m_2$ sinks inside
$r\simless 0.01$, where $M(r)\ll m_1$ and where the models have different
anisotropies. 

%++++++++++++++++++++++++++++++++++++++++++++++++++++++++++++++++++++++++++++
%++++++++++++++++++++++++++++++++++++++++++++++++++++++++++++++++++++++++++++

%++++++++++++++++++++++++++++++++++++++++++++++++++++++++++++++++++++++++++++
\begin{figure*}[tb]
\centerline{\psfig{figure=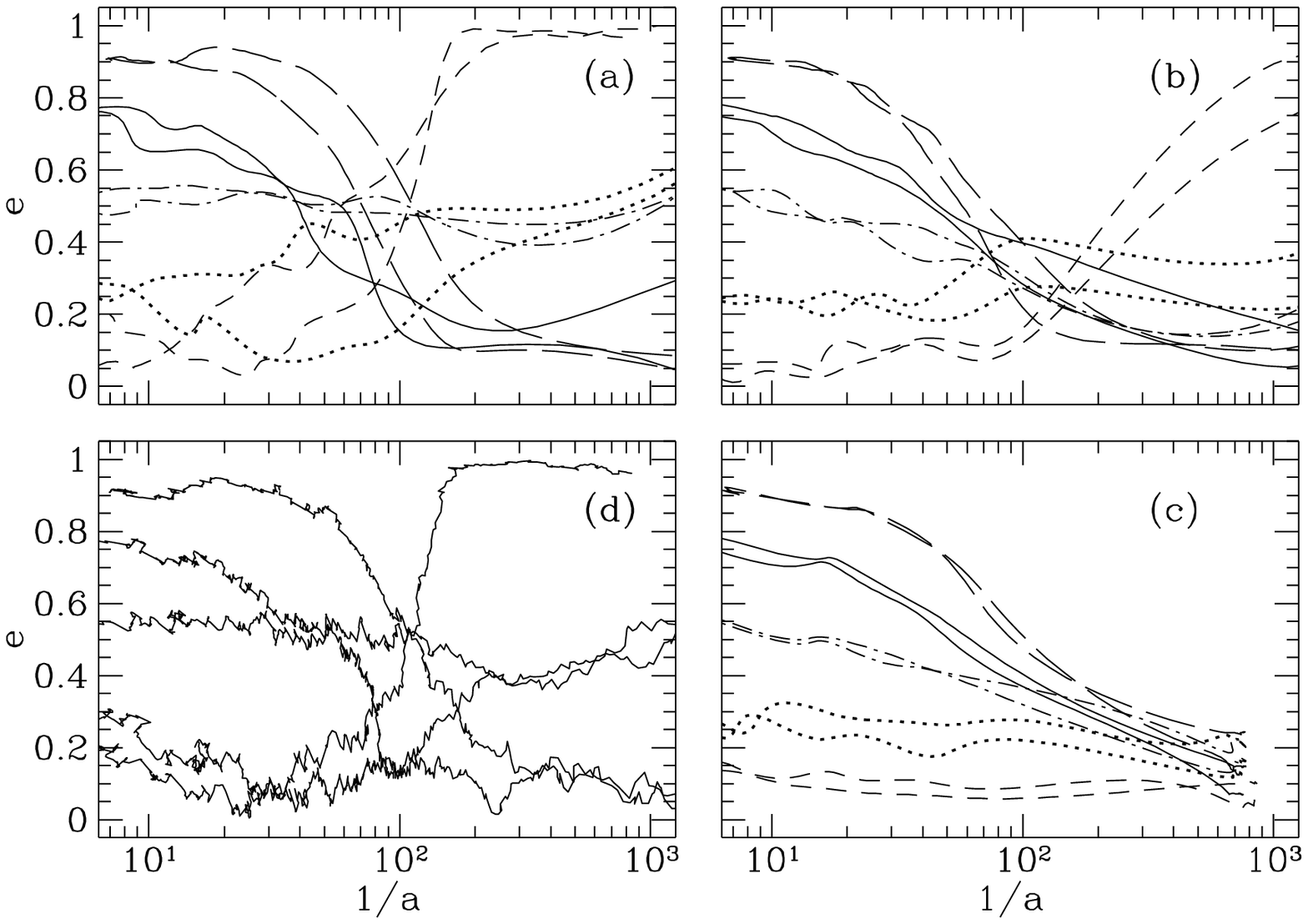,width=\the\hsize,clip=}}
\caption[Eccentricity evolution with resonant friction.]{Eccentricity versus
semimajor axis for \BHs\ of mass $m_2=0.001$ (a), 0.003 (b), and 0.01 (c)
orbiting in the five models of Figure~\ref{fig-rfrmod} (the different line
types correspond to those used in Fig~\ref{fig-rfrmod}).  The two lines for
each model and $m_2$ value are from integrations in which $m_2$ was started
in a clockwise and counter-clockwise sense.  Panel~(d) shows five of the
lines from panel~(a) before they were smoothed. The orbital elements are
defined (for this figure only) by $1/a=-2E/m_1$ and $e=\sqrt{1-L^2/L_c^2}$,
with $E$ and $L$ the energy and angular momentum of $m_2$ and with the
potential of the galaxy set to zero at $r=0.2$ for the calculation of $E$.}
\label{fig-rfreva}
\end{figure*}
%++++++++++++++++++++++++++++++++++++++++++++++++++++++++++++++++++++++++++++

The orbits of small \BHs\ were integrated in the five models starting at
$r=0.2$ and continuing until the binary semimajor axis reached 
$a\simless0.001$.  
To reduce the cpu time, $m_1$ was fixed at the origin, which is what Rauch
and Tremaine assumed.
This assumption is not valid near the end of the integrations, when the
binary is hard and the motion of $m_1$ would be important, but it does not 
affect the eccentricity much when $m_2$ first enters the region
$r\simless0.01$; we checked that by repeating a few of the integrations
without fixing $m_1$. 
Figure~\ref{fig-rfreva} shows the eccentricity evolution for $m_2$ values 
0.001, 0.003, and 0.01.
The eccentricity and semimajor axis in this figure are generalized orbital
elements that agree with the Kepler elements when the \BHs\ are close and give
useful information when they are not close (the semimajor axis then gives
an estimate of the distance between the \BHs).
Since resonant friction should increase the eccentricity if $\beta>0$ and
decrease it if $\beta<0$, the initial eccentricities were chosen to be small
for the radial models ($\beta>0$) and large for the tangential models
($\beta<0$).

Consider first the results in panel~(a) from the integrations with
$m_2=0.001$. 
The influence of anisotropy on the eccentricity is clear: in
models A and B, with radial anisotropies, the eccentricity grows larger,
whereas in models D and E, with tangential anisotropies, it grows smaller. 
In models A and E the eccentricity changes much faster than the semimajor
axis when $1/a\simeq100$, as Rauch and Tremaine predicted.
The evidence from the integrations with larger $m_2$ values is not as clear.
The eccentricity still grows in model A when $m_2=0.003$, suggesting that
resonant friction still operates, but the growth is not as fast as with
$m_2=0.001$.
When $m_2=0.01$ the eccentricity goes down in all the models, and the rate
of change is low enough to be explained by ordinary, non-resonant friction.
This $m_2$ value perturbs the resonances responsible for resonant friction.

Thus resonant friction can cause a rapid growth in the eccentricity of a \BH\
binary, but only if two conditions are satisfied: first, $m_2$ must be at
least 30 times smaller than $m_1$ (this was implicit in the formulas of Rauch
and Tremaine); second, and more troublesome, the cluster of stars bound to
$m_1$ must have a strong radial anisotropy, say $\beta>0.5$.
There is no reason to expect an anisotropy like that around a massive \BH.
The only existing models for the formation of \BHs\ in galaxies that
make definite predictions for the anisotropy---the adiabatic-growth model
(Young, 1980; Goodman and Binney, 1984; Quinlan et al., 1995), and the
binary-merger model studied here---predict a tangential anisotropy. 
Any model in which the \BH\ grows by consuming stars on radial orbits
is likely to favor a tangential anisotropy.
Some formation models may allow a radial anisotropy (e.g., models that
involve a sudden collapse at the center, with the \BH\ and the stars forming
simultaneously), but they have not been studied and are unlikely to predict
an anisotropy as strong as $\beta>0.5$.
There are two galaxies for which we have estimates of the anisotropy around
the central \BH.
If the \BH\ mass estimate of Harms et al.\ (1994) for M87 is correct, then the
best fit to the stellar velocity distribution has a tangential anisotropy,
although the error bars are large enough to allow a slight radial anisotropy
(Dressler and Richstone, 1990; Merritt and Oh, 1996).
For M32 also the best fit to the velocity distribution around the central \BH\
has a tangential anisotropy (van der Marel et al., 1997).

Although we do not believe that resonant friction will be important for the
eccentricity evolution of \BH\ binaries, the possibility of this is
nevertheless interesting: it shows the danger of relying on the simple
dynamical-friction formula. Resonant friction may be more important for
binaries in galaxies with disks, where it will affect the inclination as
well as the eccentricity.

%=============================================================================
\section{The response of the galaxy}
%=============================================================================

The response of a galaxy to the hardening of a \BH\ binary may explain why
large elliptical galaxies have weaker density cusps than smaller galaxies.
Figures \ref{fig-evjaf} and \ref{fig-mejj} showed some results for the core
expansion and mass ejection caused by a binary at the center of a Jaffe
model.  Here we examine the change in the density profile for a range of
binary masses, and also the kinematical signature that the binary leaves in
the velocity distribution.  We again use a Jaffe model to represent the
galaxy; the results will apply with some changes to models with other gamma
values.

\begin{figure*}[tb]
\centerline{\psfig{figure=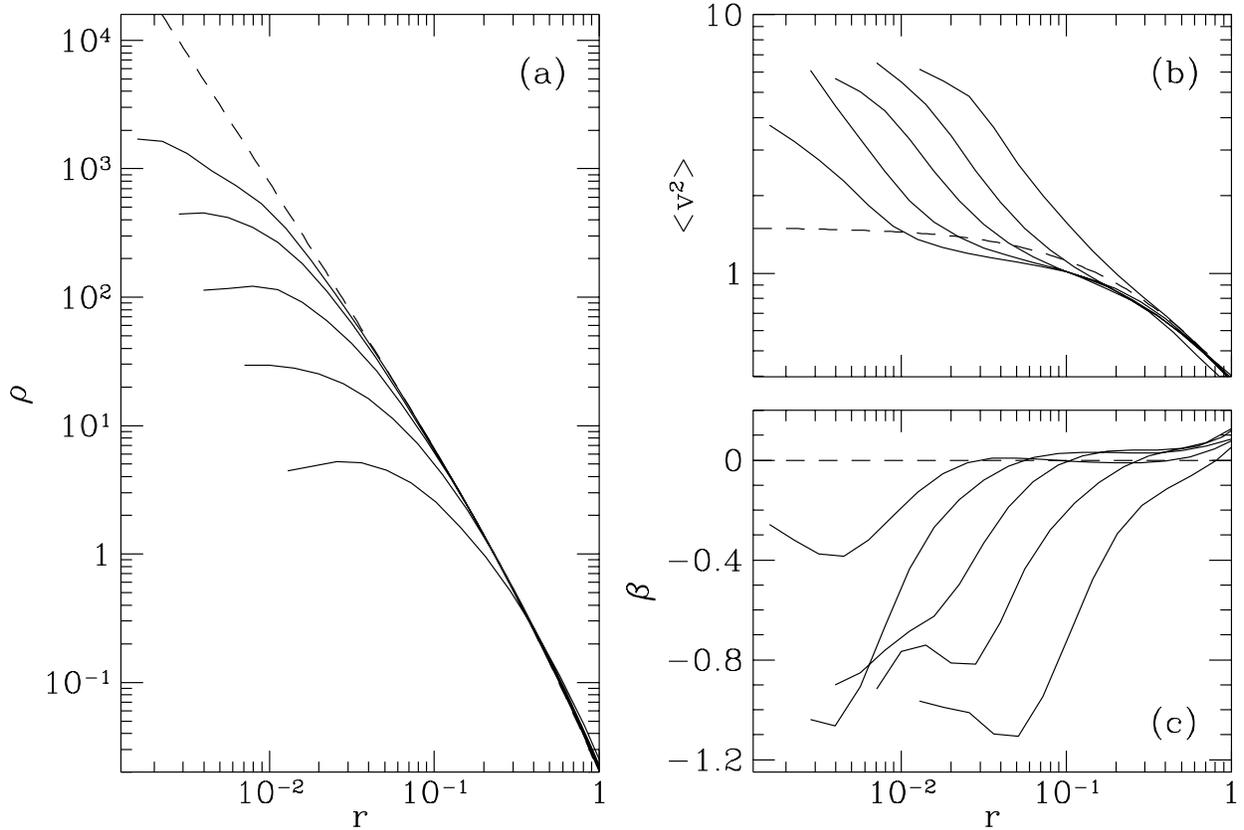,width=\textwidth,clip=}}
\caption[Response of the galaxy.]{Response of a Jaffe model to the hardening
of a \BH\ binary.  The dashed lines show the initial (a) density, (b) 3D
velocity dispersion, and (c) anisotropy.  The solid lines show these
quantities when the binary reaches $\Vbin=4$, with the binary started as in
Fig.~\ref{fig-evjaf}, and with the five lines for \BH\ masses $m_1=m_2=0.04$
(rightmost line), 0.02, 0.01, 0.005, and 0.0025 (leftmost line).}
\label{fig-jgalr} \end{figure*}

We integrated binaries with five masses ($m_1=m_2=0.04$, 0.02, 0.01, 0.005,
and 0.0025), starting them as we did for the integration in
Figure~\ref{fig-evjaf}.  The galaxy model had $10^5$ particles with a mass
exponent $\lambda=0.75$.  To study the change in the galaxy we kept frequent
records of the radii containing fixed fractions of the total mass, like the
radii in panel~(c) of Figure~\ref{fig-evjaf} but more of them, and also of
the sums of $m v_r^2$ and $m v_t^2$ for the stars inside those radii.  By
differentiating the mass fractions and velocity sums with respect to radius,
after first smoothing their time dependence, we were able to recover the
density and the radial and tangential dispersions at any time during the
integrations.  We use this data in Figure~\ref{fig-jgalr} to plot the
densities and dispersions when the five binaries reached $\Vbin=4$, the
highest velocity reached by the binary with $m_1=0.04$ (we stopped the
integration then because it was becoming slow).
%++++++++++++++++++++++++++++++++++++++++++++++++++++++++++++++++++++++++++++
%++++++++++++++++++++++++++++++++++++++++++++++++++++++++++++++++++++++++++++
The binaries with smaller masses reached higher velocities, but to study the
dependence on the binary mass it is best to compare the galaxies at the same
$\Vbin$.

The density profiles in panel~(a) show that a binary can eliminate a density
cusp.  Although the profile for the smallest binary has a mild cusp, the
others have densities that are flat or, in some cases, decreasing towards
the center; they would probably all decrease if we could plot them closer to
the center (this was predicted by Begelman et al.\ 1980).  The mass ejected
from the central region is $\Mej/\M12=1.2$--1.5, as we reported in
Figure~\ref{fig-mejj}, with no systematic dependence of $\Mej/\M12$ on the
binary mass.  Because the Jaffe model has a mass that, near the center,
increases linearly with radius, the core radius at a fixed $\Vbin$ increases
linearly with the binary mass; for other gamma models the dependence would
be nonlinear.  The $\Mej/\M12$ values computed here are larger than we
predicted in paper~I, partly because some mass gets displaced from the
center in the N-body experiments before the binaries form and partly
because, as we discussed in Section~\ref{sec-wand}, the ejection rates once
the binaries form in the N-body experiments (and, we believe, in real
galaxies) are higher than we predicted in paper~I.  As the binaries continue
hardening, more mass will be ejected and the core radii will grow.  For
example, Figure~\ref{fig-mejj} shows the density profile for the binary with
$m_1=0.01$ at $\Vbin=6.8$: the central density is less than half of its
value at $\Vbin=4$, and the core radius is nearly twice as large.  The final
profile will depend on how far the binary has to harden before gravitational
radiation becomes important.  Massive \BH\ binaries in typical galaxy cores
have to shrink by a factor of 10--50 beyond the point at which they become
hard, with the factor varying with the \BH\ masses as $\M12^{2/5}$ for an
equal-mass binary.  The binaries in Figure~\ref{fig-jgalr} shrank about a
factor of 4 beyond the point at which they became hard.  In a typical galaxy
they would have to shrink another factor of 3--12, causing $\Vbin$ to rise
by a factor of 1.7--3.5 and, using Figure~\ref{fig-mejj} as a guide, adding
0.4--0.8 to the value of $\Mej/\M12$.

The kinematical response of the galaxy is shown in panels (b) and (c).  The
velocity dispersions rise towards the center, approximately (but not
exactly) as $v^2\sim 1/r$.  Mass estimates derived from these dispersions
will underestimate the central masses if the anisotropies are not taken into
account.  The models all develop strong tangential anisotropies, because the
stars on radial orbits are the ones that get ejected.  Four of the five
lines in panel~(c) show $\beta\simeq-1$ near the center, although we see no
reason why that value should be preferred.  The central anisotropy is
difficult to measure (the leftmost parts of our lines are affected by
noise); we expect that $\beta$ would fall below $-1$ if we could measure it
closer to the binary.  This is a much stronger anisotropy than is predicted
by models like the adiabatic-growth model for the formation of massive black
holes. The adiabatic growth of a \BH\ in an initially isotropic galaxy leads
to a final anisotropy of at most $\beta=-0.3$ (Quinlan et al., 1995)

Our mass-ejection results support the suggestion of Ebisuzaki et al.~(1991)
and Faber et al.~(1996) that \BH\ binaries can explain the weak density
cusps in large elliptical galaxies.  If these galaxies contain massive
\BHs, and if they have formed through mergers of smaller galaxies also
containing massive \BHs, the binaries that result will clear out the central
region.  A real galaxy merger will of course be more complicated than the
idealized experiments considered here.  In most cases a binary will result
when a small satellite galaxy sinks into a large galaxy containing a massive
\BH.  The scattering experiments from paper~I suggested that more mass gets
ejected if a massive \BH\ merges with a number of small \BHs\ than if it merges
with one large \BH\ (with the same total mass), although each small \BH\ may
counteract the mass ejection to some extent by bringing fresh stars and gas
to the center.  The general result from our experiments should remain true:
the density cusp in a merged galaxy will be weaker if the merger leads to a
\BH\ binary.  And with this result we can make a testable prediction: if this
is the explanation for the weak density cusps in large galaxies, the central
velocity distributions in those galaxies should have strong tangential
anisotropies.  As we mentioned in Section~\ref{sec-ecc}, there is evidence
that the galaxy M87 passes this test.

%==============================================================================
\section{The binary merger time in real galaxies}
%==============================================================================

The answer to many questions involving massive \BH\ binaries depends on the
binary merger time.  A big uncertainty in computing this has been loss-cone
depletion.  Although the loss cone can be replenished if the potential is
nonspherical, and the binary merger can be helped by gas accretion onto the
\BHs, for many galaxies these perturbations will not be enough.  We therefore
consider whether a binary can merge through purely stellar dynamical
processes in a spherical galaxy.  We believe that the answer is often yes,
thanks to the wandering of the binary from the galaxy center.

Previous discussions of the loss-cone problem have mostly assumed that the
central object (a single \BH\ or a binary) remains fixed at $r=0$.  The rate
at which stars accrete onto or interact with the object once the loss cone
becomes depleted is then given by an expression of the form (e.g., Shapiro,
1985)
\begin{equation}                                                \label{eq-fuel}
   {dM\over dt} \sim {\rho(\rcrit)\rcrit^3 \over t_r(\rcrit)
   \ln(\rcrit/\rD)},
\end{equation}
where $\rD$ is the destruction or interaction radius, which we take to be
the binary semimajor axis, and $\rcrit$ is the radius at which the loss-cone
angle $\thetalc$ equals the rms deflection suffered by stars through
two-body relaxation in one dynamical time.  For \BH\ binaries in typical
galaxies, $\rcrit$ is larger than $G\M12/\sigma^2$, and the relaxation time
at $\rcrit$ is much longer than the age of the galaxy.  The hardening rate
implied by equation~(\ref{eq-fuel}) is then too low to allow most binaries
to merge.

Valtonen (1996) used the galaxy M87 to illustrate how difficult it is for a
central binary to merge when the relaxation time is long.  He considered an
equal-mass binary with $\M12=3\times10^9\,M_\odot$ and a separation
$a=0.6\,$pc, at which the merger time from gravitational radiation is
$10^{10}\,$yr, and at which the total mass in stars that can approach within
$a$ of the center---assuming that the loss cone is not depleted---is only
1/200 of the mass of the binary.  For the binary to shrink by a factor of
$e$, the loss cone would have to be refilled about 200 times.  The problem
is worse for smaller \BHs\ because they have to shrink more before radiation
becomes important. For $\M12=3\times10^6\,M_\odot$ the factor of 200 in
Valtonen's example gets replaced by 7800.  Valtonen concluded that, for
nearly the whole range of parameters of interest, \BH\ binaries in galaxies
are incapable of merging.

Wandering changes this conclusion. We refer here not to the small correction
to the hardening rate that we discussed in Section~\ref{sec-wand}, but to a
larger correction that occurs when the loss cone becomes depleted. A simple
equipartition estimate predicts that a particle of mass $\M12$ (a single
\BH\ or a binary) in a homogeneous core of smaller particles of mass $m$
will have a maximum wandering amplitude
\begin{equation}                                                  \label{eq-rw}
   \rw \simeq \rc\left( m\over \M12\right)^{1/2}.
\end{equation}
Young~(1977) argued that this amplitude is sometimes large enough to affect
the growth of \BHs\ in the centers of galaxies, but most subsequent work has
either ignored wandering or has dismissed it as unimportant.  For \BH\
binaries we believe that wandering is even more important than Young
suggested, because the prediction~(\ref{eq-rw}) underestimates the wandering
amplitude of a \BH\ binary. The prediction is applicable to a massive
particle in a homogeneous core, but a \BH\ binary ejects stars from the
center of the core, reducing the restoring force that keeps it at the
center, and the superelastic encounters between the binary and the stars
invalidate the equipartition assumption on which the prediction is based.
Consider the integration that we used earlier as an example.  The core
radius in Figure~\ref{fig-mejj} is $\rc\simeq0.04$ (derived by fitting a
King-model core to the final density), and the mass ratio is $m/\M12\simless
10^{-5}/0.02$ (this would be an equality if the stars had equal masses), so
the prediction~(\ref{eq-rw}) gives $\rw\simless0.0009$. But the maximum
wandering amplitude shown in Figure \ref{fig-evjaf} is more than five times
as large. And, as we mentioned in Section~\ref{sec-sample}, the amplitude
remained the same when we raised $N$ from $10^5$ to $2\times 10^5$, making
the discrepancy with the prediction even larger.

We suspect that in many galaxies it is the wandering of the binary, and not
diffusion of stars into the loss cone, that determines the hardening rate
once the loss cone gets depleted. In our Jaffe-model integration, for
example, the hardening rate did not increase when we added non-spherical
terms to the potential by changing $\lmax$ from 0 to 2, even though this
change made the potential noiser, somewhat like the way that two-body
relaxation does, and allowed more stars to reach the center.  The change did
increase the hardening rate when we applied a constraint force to fix the
binary at the center of the galaxy.

Wandering will not be enough to allow all \BH\ binaries to merge.  It allows
a binary to continue hardening beyond the point at which it would stop
without wandering, but the hardening continues at a reduced rate.  As a
rough guide, we can say from our integrations that once a binary ejects the
stars from its immediate vicinity and creates a core, it hardens at a rate
that is about 10--50 times slower than the rate we predicted in paper~I
ignoring loss-cone depletion.  Whether this will be fast enough to allow a
merger depends on the particular galaxy under study. Gas and nonspherical
perturbations will help if they are present.

These conclusions have been derived for equal-mass binaries.  The most
likely formation scenario for a massive \BH\ binary is for a small satellite
galaxy containing a central \BH\ to sink into a larger galaxy with a larger
\BH.  The bottleneck in this case may be the time required for the small
\BH\ to reach the center.  N-body experiments by Balcells and Quinn (1989,
1990) show that small, dense satellites can easily sink into the center of
large elliptical galaxies, but Weinberg's (1997) perturbation calculations
suggest that this is not always the case.  If the satellite galaxy
surrounding the small \BH\ gets disrupted, the \BH\ will take longer to
reach the center.  Neither Weinberg nor Balcells and Quinn included \BHs\ in
their models; some preliminary experiments with \BHs\ have been done by
Governato et al.~(1994).  More work on this is needed, since the predicted
merger rate for the gravitational-wave detector LISA is much larger if small
\BHs\ can participate in the mergers (Haehnelt, 1994).

%=============================================================================
\section*{Acknowledgements}
%=============================================================================

We thank Sandra Faber, Jeremiah Ostriker, Martin Rees, and Scott Tremaine
for helpful discussions while this work was in progress. We are especially
grateful to Sverre Aarseth for helping us use his N-body integration
routines in our hybrid program.  Financial support was received at UCSC from
the NSF under grant ASC 93-18185 and the Presidential Faculty Fellows
Program, and at Rutgers from NSF grant AST 93-18617 and NASA Theory grant
NAG 5-2803.

\appendix

%==============================================================================
\section{The hybrid N-body program}
%==============================================================================

We describe here some technical details of our hybrid N-body program.  We
assume that the star cluster contains two \BHs, but the program can work with
any number of \BHs.  Our notation follows that of Aarseth (1994) and
Hernquist and Ostriker (1992).

%------------------------------------------------------------------------------
\subsection{The integrator}
%------------------------------------------------------------------------------

The program uses different integrators for the \BHs\ and the stars.
The stars are integrated with the NBODY1 integrator.
The stepsizes for the stars vary continuously and are updated after each
step by a function involving the force and the first three time
derivatives of the force, multiplied by the square root of an accuracy
parameter $\eta$.
We used $\eta=0.005$--0.01, smaller than the value that Aarseth recommends
($\eta=0.02$--0.03), so that we could integrate close encounters accurately
with a small softening length. 
The \BHs\ are integrated with the NBODY2 integrator, which uses the
Ahmad-Cohen neighbor scheme.
The \BHs\ have two stepsizes, $\delta t_i$ and $\delta t_r$, determined by two
accuracy parameters, $\eta_i=\eta$ and $\eta_r=2\eta_i$. The \BHs\ are moved
and the irregular force from nearby stars (the neighbors) is updated on
the smaller stepsize $\delta t_i$; the regular force from the remaining
stars is updated on the larger stepsize $\delta t_r$ and is predicted
between those times from its known time derivatives.
The number of neighbors for each \BH\ is set to approximately $(N/4)^{3/4}$,
the scaling recommended by Makino and Hut (1988); the program forces each \BH\
to be in the neighbor list of the other \BH\ at all times.
When $N$ is large, $\delta t_i$ is usually much smaller than $\delta t_r$.
The neighbor scheme improved the efficiency of our integrations by at least
a factor of ten. 

The program checks periodically to see if the \BHs\ are close enough for their
motion to be regularized.
If they are, the \BHs\ are replaced by a center-of-mass particle and a set of
KS coordinates describing their relative motion.
Our criterion for ``close enough'' is that each \BH\ is the closest neighbor
of the other \BH, that the force on each \BH\ from the other \BH\ is five times
larger than the force from the stars, and that the neighbor list for the
center-of-mass particle is large enough to contain the perturbers for the KS
coordinates, which means the stars that exert a dimensionless perturbation
exceeding a value $\gmin$.
The KS equations of motion are integrated with the Hermite integrator from
the NBODY6 program; the center-of-mass motion is integrated with the NBODY2
integrator as is the motion of single \BHs, except that the forces between
the center of mass and nearby stars (those within a distance $\lambda a$)
are computed by resolving the binary into its two \BHs.
We used $\lambda=1000$, probably much larger than was necessary (Aarseth
recommends $\lambda=70$).
The KS stepsize is chosen so that the number of steps per unperturbed
period is $2\pi/\eta_u$, with $\eta_u$ the KS accuracy parameter
(Aarseth recommends $\eta_u\simless0.1$; we used 0.07).
If the binary has no perturbers, the Keplerian motion is integrated
exactly. Even with perturbers, regularization allows close binaries to
be integrated efficiently regardless of their eccentricity, with no
softening needed for the \BH-\BH\ interaction.

The large difference between the masses of the \BHs\ and the stars in our
experiments caused some difficulties with the standard KS integration
procedures. 
When we used the value for $\gmin$ that Aarseth recommends ($10^{-6}$) the
stepsize for the center-of-mass particle was sometimes much smaller than the
stepsize for the KS coordinates, because a small star near the massive
binary does not get included in the list of KS perturbers if its
dimensionless perturbation is less than $\gmin$, but it does get included in
the center-of-mass neighbor list and it affects the center-of-mass stepsize.
Large errors can result if there are many such stars.
To improve the accuracy we reduced $\gmin$ to $10^{-7}$, and we introduced a
parameter $\Gamma=50$ so that stars within a distance $\Gamma a$ of the
binary are included in the KS perturber list whatever the size of their 
perturbation.
Although these precautions slowed our integrations somewhat, the integration
of close binaries still went several times faster when regularization was
used.

Because the program regularizes only the motion of the \BH\ binary, and not
the motion of stars approaching the binary, the interactions between the \BHs\
and the stars have to be softened; otherwise some stars would be given tiny
stepsizes when they encounter the \BHs.
We chose the softening length so that the softening velocity
(eq.~\ref{eq-vsoft}) was at least 2--3 times as large as the velocity
of the binary at the end of the integration; three-body scattering
experiments showed that a softening length of that size does not affect
the average energy transfer.
Even with softening we sometimes had trouble with stars that got captured
into tightly bound orbits around the \BHs. 
We therefore merged stars with the \BHs\ if they were captured into orbits
with semimajor axes less than twice the softening length. 
The number of stars merged in this way was small: the growth of the \BH\ masses
was at most 0.2\%.

%------------------------------------------------------------------------------
\subsection{The expansion of the potential}
%------------------------------------------------------------------------------

The coefficients $A_{nlm}$ for the potential expansion are updated at
fixed time intervals $\Delta t$, after the coordinates of the stars are first
predicted to a common time.
We used the basis functions of Hernquist and Ostriker (the lowest-order
member of the set is the Hernquist potential), but adjusted the scale length
$d$ for each initial model to customize the fit (Hernquist and Ostriker
assume that $d=1$). 
We found it necessary to soften the basis functions by replacing the radius
$r$ of a particle by $(r^2+h^2)^{1/2}$, with $h\simeq 10^{-4}$, for
otherwise our use of both a mass spectrum and an integrator with individual
stepsizes sometimes caused stars to get trapped in orbits with tiny
stepsizes at the center of the galaxy. 
In most of our integrations we used a spherical basis ($\lmax=0$), with
the number of functions in the range $10\leq\nmax\leq20$, depending on
the initial model (the Jaffe model required largest $\nmax$).
With a spherical basis the expansion assumes that the center of
mass of the galaxy is fixed at the origin, an assumption that has been shown
to artificially accelerate the orbital decay of satellite galaxies in the
sinking-satellite problem (White, 1983), although only when the satellite is
far from the center of the primary galaxy (Quinn and Goodman, 1996).
The assumption should not affect the evolution rate in our \BH-binary
experiments, because the \BHs\ start close to the center and often
symmetrically about the center, making the dipole distortion of the galaxy
small.
When we used a spherical basis the cpu time spent computing the expansion
functions was at most half of the total cpu time; most of the time was spent
computing the forces between the \BHs\ and the stars.

The NBODY1 integrator needs the forces and the first three time derivatives
of the forces acting on the stars at the start of an integration.
This initialization is cumbersome when the expansion method is used.
In principle the derivatives could be computed analytically from the
expansion functions, but in practice we used the known derivatives for the
initial galaxy model, or used numerical interpolation to approximate the
derivatives.
This would have been simpler if we had used the Hermite integrator for the
stars, as that requires only the first time derivative (Makino 1991).

%------------------------------------------------------------------------------
\subsection{Tests and possible improvements}
%------------------------------------------------------------------------------

We ran a number of tests to check the integrators and the potential
expansion.
We first fixed the expansion coefficients and checked that the conservation
of energy improved in the manner expected for Aarseth's fourth-order
integrators when we reduced the accuracy parameters $\eta$ and $\eta_u$.
We also checked that the evolution of a hard binary was independent of which
integrator we used for the \BHs\ (NBODY1, NBODY2, or NBODY6), provided that
the accuracy parameters were small (the required cpu time, of course,
depends on the integrator).
To test the potential expansion we checked that the program could maintain
equilibrium $\gamma$ models without \BHs\ for radii $r\simgreat 10^{-2}$ (the
minimum radius depends on $\gamma$ and $\nmax$).
We also repeated some of our \BH-binary experiments using the basis functions
of Clutton-Brock (1973) instead of those of Hernquist and Ostriker, to check
that the two gave the same result when $\nmax$ was large.

One test did reveal an important limitation of our program.
We ran some cold-collapse experiments (starting a galaxy with the stars at
rest, with no \BHs) to check that the SCFBDY and SCF programs gave the same
result for the virial ratio versus time.
They did, but only if we used a stepsize $\Delta t$ with SCFBDY considerably
smaller than the stepsize required with SCF.
The reason is that with SCFBDY the updating of the expansion coefficients is
not time reversible.
The energy error after a fixed time therefore varies linearly with $\Delta
t$ with SCFBDY, and not quadratically as it does with SCF. 
This did not matter for our work, because the change in the potential was
slow and we were able to choose $\Delta t$ small enough to make the errors
small, but it would matter for a problem with a rapidly changing potential.

Some improvements to the program may be possible.
The particles could be given block stepsizes to make the program run faster
on vector computers, although the vectorization will always be hindered by
the complicated integration algorithm.
The need to soften the \BH-star interactions could be removed by
more-sophisticated regularization, although this would be difficult when 
many stars interact with the \BHs\ simultaneously.
And the limitations of the potential expansion---such as the linear
dependence of the error on $\Delta t$, and the difficulty of resolving small
stellar clumps away from the center---could be reduced by updating the
contribution to the expansion coefficients from the inner part of the galaxy
more frequently than the contribution from the outer part, or by using the
direct-summation method for the inner part and an expansion method only for
the outer part, although either of these changes would increase the program
complexity, and using the direct-summation method would increase the cpu
time. 
We did not try any of these ideas.

%=============================================================================

%============================================================================

\end{document}